\documentclass[12pt,preprint]{aastex}
\slugcomment{to appear in the Astrophysical Journal Letters}

\shorttitle{CO emission at $z = 5.77$}
\shortauthors{Carilli et al. }

\begin{document}

\title{Detection of $1.6\times 10^{10}$ M$_\odot$ of molecular gas in \\ 
the host galaxy of the $z=5.77$ SDSS quasar J0927+2001.}

%% Use \author, \affil, and the \and command to format
%% author and affiliation information.

\author{C.L. Carilli\altaffilmark{1},
R. Neri\altaffilmark{2},
R. Wang\altaffilmark{1,13}, 
P. Cox\altaffilmark{2},
F. Bertoldi\altaffilmark{3},
F. Walter\altaffilmark{4},
X. Fan\altaffilmark{5},
K. Menten\altaffilmark{6},
J. Wagg\altaffilmark{1},
R. Maiolino\altaffilmark{7},
A. Omont\altaffilmark{8},
Michael A. Strauss\altaffilmark{9},
D. Riechers\altaffilmark{4},
K.Y. Lo\altaffilmark{10},
A. Bolatto\altaffilmark{11},
N. Scoville\altaffilmark{12}
}

\altaffiltext{1}{National Radio Astronomy Observatory, PO Box O, 
Socorro, NM, USA, 87801, ccarilli@nrao.edu}
\altaffiltext{2}{Institut de Radio Astronomie Millimetrique (IRAM), 300 
rue de la Piscine, Domaine Universitaire de Grenoble, 38406 
St. Martin d'Hères, France}
\altaffiltext{3}{Argelander-Institut fur Radioastronomie, Universitat Bonn,
auf dem Hugel 71, Bonn, 53121, Germany}
\altaffiltext{4}{Max-Planck Institut f\"ur Astronomie, Konigstuhl 17,
Heidelberg, Germany}
\altaffiltext{5}{(Steward Observatory, The University of Arizona, 
Tucson, AZ 85721}
\altaffiltext{6}{Max-Planck Institut f\"ur Radioastronomie, Auf dem H\"ugel
69, Bonn, 53121, Germany}
\altaffiltext{7}{Osservatorio Astrofisico di Arcetri, Largo E. Fermi 5, 
50125 Firenze, Italy}
\altaffiltext{8}{Institut d'Astrophysique de Paris, CNRS; and 
Universite Pierre et Marie Curie, Paris, France}
\altaffiltext{9}{Princeton University Observatory, Princeton, NJ 08544, USA}
\altaffiltext{10}{National Radio Astronomy Observatory, 520 Edgemont 
Road, Charlottesville, VA, USA}
\altaffiltext{11}{Department of Astronomy, 601 Campbell Hall, 
University of California at Berkeley, CA 94720}
\altaffiltext{12}{Robinson Laboratory, California Institute of Technology, 
Pasadena, CA 91125}
\altaffiltext{13}{Astronomy Department, Peking University, 
Beijing 100871, China}

\begin{abstract}

We have detected emission by the CO 5-4 and 6-5 rotational transitions
at $z = 5.7722\pm 0.0006$ from the host galaxy of the SDSS quasar
J0927+2001 using the Plateau de Bure interferometer. The peak line
flux density for the CO 5-4 line is $0.72 \pm 0.09$ mJy, with a line
FWHM = $610 \pm 110$ km s$^{-1}$. The implied molecular gas mass is
$(1.6 \pm 0.3) \times 10^{10}$ M$_\odot$. We also detect the 90 GHz
continuum at $0.12 \pm 0.03$ mJy, consistent with a 47K dust spectrum
extrapolated from higher frequencies.  J0927+2001 is the second
example of a huge molecular gas reservoir within the host galaxy of a
quasar within 1 Gyr of the big bang.  Observations of J0927+2001 are
consistent with a massive starburst coeval with a bright quasar phase
in the galaxy, suggesting the rapid formation of both a super-massive
black hole through accretion, and the stellar host spheroid, at a time
close to the end of cosmic reionization.

\end{abstract}


\section{Introduction}

Understanding the relationship between quasars and their host galaxies
has become imperative since the discovery of the bulge mass -- black
hole mass correlation in nearby galaxies, a result which implies a
fundamental relationship between the formation of supermassive black
holes (SMBH) and spheroidal galaxies (Ferrarese \& Merrit 2000;
Magorrian et al. 1998; Gebhardt et al. 2000).  In our extensive study
of the dust and gas content, and star formation activity, of the host
galaxies of $z > 2$, luminous ($M_{1450\AA} < -26.1$), optically
selected high redshift quasars (Omont et al. 2003; Carilli et al. 2002;
Bertoldi et al. 2003a; Petric et al.  2003; Beelen et al. 2006; Wang
et al. 2007a), we have found that roughly 30\% of these sources are
also hyper-luminous infrared galaxies ($L_{FIR} \sim 10^{13}$
L$_\odot$). This FIR emission corresponds to thermal emission from
warm dust. Searches for CO emission from these sources typically yield
molecular gas masses $\ge 10^{10}$ M$_\odot$ (Solomon \& Vanden Bout
2005).  Such molecular gas represents the fuel for star formation, and
can serve as a dynamical tracer in a forming galaxy.

An important finding of our program is that the 30\% fraction of FIR
luminous quasars continues to the highest redshifts, $z \sim 6$,
approaching the end of cosmic reionization, and hence pushing toward
the very first galaxies and SMBHs (eg. Fan, Carilli, \& Keating 2006;
Fan et al.  2006a).  A particularly enlightening example is the
highest redshift quasar known, J1148+5251 at $z=6.419$. The host
galaxy has been detected in thermal dust, non-thermal radio continuum,
and CO line emission (Walter et al. 2003; Bertoldi al. 2003b; Carilli
et al. 2004), with an FIR luminosity (integrated between $42.5 \mu$m
and 122$\mu$m) of $1.3\times 10^{13}$ L$_\odot$ (Beelen et al. 2006),
a molecular gas mass of $\sim 2\times 10^{10}$ M$_\odot$, and a dust
mass of $7\times 10^8$ M$_\odot$.  The dynamical mass to a radius of
2.5 kpc in the host galaxy, as estimated from high resolution VLA
imaging of the molecular gas distribution, is comparable to the gas
mass, but much less than the mass predicted by the black hole mass --
bulge mass relation measured at low redshift (Walter et al. 2004).  We
have also detected the dominant cooling line of interstellar gas,
[CII] 158 $\mu$m, from J1148+5251 (Maiolino et al. 2005). These
observations demonstrate that large reservoirs of dust and metal
enriched atomic and molecular gas can exist in massive galaxies within
1 Gyr of the Big Bang.  The current observations suggest active star
formation in the host galaxy, with a massive star formation rate of
order $10^3$ M$_\odot$ year$^{-1}$, although the question remains as
to the contribution to dust heating by the AGN (Wang et al. 2007a; Li
et al. 2007a).

The source SDSS J092721.82+200123.7 (hereafter J0927+2001) is similar
to J1148+5251 in many ways.  J0927+2001 is a quasar selected from the
Sloan Digital Sky Survey (York et al.  2000), with an optical redshift
of $z =5.77 \pm 0.03$ (corresponding to $\rm t_{univ} = 1$ Gyr,
assuming a standard concordance cosmology), based on fitting to the
(relatively weak) Ly$\alpha$ + NV lines, and with L$_{bol} = 7\times
10^{13}$ erg s$^{-1}$ (Fan et al. 2006b; Wang et al. 2007a).  The black
hole mass estimated from both the Eddington limit, and from UV line
widths, is $\sim 10^9$ M$_\odot$ (Jiang et al. 2007).

J0927+2001 has been detected in 250 GHz continuum emission at the IRAM
30m telescope, with a flux density of $5.0\pm 0.8$ mJy (Wang et
al. 2006), and at 350 GHz at the Caltech Submm Observatory, with a
flux density of $18\pm 5$ mJy (Wang et al. 2007b, in
preparation). These data, in combination with lower and higher
frequency data, reveal a clear FIR excess over a standard quasar IR
SED. This FIR excess corresponds to thermal emission from warm dust
with a temperature $\sim 47$ K.  The dust mass is $6.9\times 10^8$
M$_\odot$, and the FIR luminosity is $1.2\times 10^{13}$ L$_\odot$.  A
weak ($45\pm 14\mu$Jy) radio continuum counterpart has been detected
at 1.4 GHz, and the rest-frame radio through FIR SED is within the
range defined by star forming galaxies (Wang et al. 2007a). The
observations of J0927+2001 are consistent with a massive (stars $> 5$
M$_\odot$) star formation rate of $\sim 700$ M$_\odot$ year$^{-1}$
(Kennicutt 1998), or a total star formation rate (stars $> 0.1$
M$_\odot$) a factor 5.6 higher, assuming a Salpeter initial mass
function, and assuming the warm (47 K) dust is heated by star
formation.

In this letter we present the discovery of CO 5-4 and 6-5 emission
from J0927+2001.  These observations reveal the necessary fuel for
star formation, and enable a number of key observations in the
study of the earliest generation of massive galaxies and SMBH. 

\section{Observations and results}

Our observations have been made with the improved Plateau de Bure
interferometer equipped with the new generation of receivers. With
their increased sensitivity (detection of sub-mJy lines in a single
track), and increased bandwidth (2 of the 4 GHz with the current
correlator), these new receivers allow for deep searches for faint
redshifted molecular emission lines over bandwidths adequate to cover
the typical uncertainties of optical redshifts of high redshift
quasars.

We searched for CO 5-4 emission from J0927+2001 with the Plateau de
Bure Interferometer in April and May, 2007.  The first track (8 hours)
used two polarizations, and two frequency settings which covered a
total bandwidth of 1.8 GHz (= 5700 km s$^{-1}$ or $\Delta z$ = 0.13)
centered at $z=5.79$ (or 84.87 GHz for redshifted CO 5-4). After a
possible line was identified in the first observations at $z \sim
5.77$, a second 8 hour track was observed centered on the line at
85.12 GHz, covering a total bandwidth of 1 GHz using dual
polarization.  This second track confirmed the 5-4 line with high
significance. A third 8 hour track was then observed centered on the
frequency of the CO 6-5 line redshifted to 102.10 GHz, which was also
clearly detected. The spectral resolution for all observations was 2.5
MHz channel$^{-1}$.  The phase stability, as measured on the phase
calibrator (0851+202) was easily adequate for coherent integration in
the D configuration, for which the interferometric synthesized beam
had a FWHM $\sim 5"$. Absolute gain calibration was performed on the
star MWC349.

Figure 1 shows the spectra of the CO 5-4 and 6-5 emission lines from
J0927+2001 at a spectral resolution of $70$ km s$^{-1}$
channel$^{-1}$. The rms per channel varies slightly across the
spectra, but typical values for the 5-4 spectrum are 0.37 mJy
beam$^{-1}$ for 5-4, and 0.52 mJy beam$^{-1}$ for 6-5.

Both CO emission lines are clearly detected, as well as the underlying 
continuum emission at both frequencies. 
Gaussian fitting to the lines yield the parameters
listed in Table 1, including: the continuum level, the line peak flux
density and FWHM, the velocity integrated line flux, and the line
luminosity, $L'(CO) = 3.25\times 10^{13} (1+z) I \Delta v D_A^2
\nu_{obs}^{-2}$ K km s$^{-1}$ pc$^{2}$, where $I \Delta v$ is the
velocity integrated flux density in Jy km s$^{-1}$, $D_A$ is the
angular diameter distance in Gpc ($D_A$ = 1.23 Gpc at $z=5.77$), and
$\nu_{obs}$ is the observing frequency in GHz (Solomon et al. 1997). A
linewidth of $600 \pm 70$ km s$^{-1}$, and an LSR line centroid of $z
= 5.7722\pm 0.0006$, are derived from the spectrum obtained by merging
the CO 5-4 and 6-5 data.

The measured continuum level for the combined data is $0.12 \pm 0.03$
mJy.  Combined with the continuum observations at 250 GHz and 350 GHz,
this yields a dust temperature of 47 K, assuming a modified grey body
of index $\beta = 1.6$ (see analysis in Wang et al. 2007b).

Figure 2 shows the image of the velocity integrated line emission from
the weighted sum of the 5-4 and 6-5 lines.  The peak of the CO
emission is at: (J2000) $09^{\rm h}27^{\rm m}21.79^{\rm s}~
+20^{\circ}01'23.5''$, or within 0.2$"$ of the optical position. The
source appears unresolved at the present resolution of FWHM $\sim 5"$.

\section{Analysis}

The ratio of the strength of the two CO transitions is: $L'_{6-5} /
L'_{5-4} = 1.10 \pm 0.27$. This ratio is consistent with constant
brightness temperature (ie. ratio = 1) up to CO 6-5. The high CO
excitation in J0927+2001 is comparable to that seen in the host galaxy
of J1148+5251 (Bertoldi et al. 2003), and other high redshift quasars
(Solomon \& Vanden Bout 2005; Carilli et al. 2002; Weiss et al.  2007
in prep), as well as in the nuclei of nearby nuclear starburst
galaxies, such as NGC 253 and Arp 220 (Greve et al. 2006; Bradford et
al. 2003), implying warm ($> 50$K), dense ($>10^4$ cm$^{-3}$)
gas. However, given the close spacing of the two transitions,
observations of a lower order transition (1-0 or 2-1) are needed to
better constrain the gas excitation conditions.

We derive the total molecular gas mass (dominated by H$_2$) from the
CO luminosity using the conversion factor for CO luminosity to total
molecular gas mass appropriate for Ultra-luminous Infrared galaxies
(ULIRGs, or galaxies with $L_{FIR} > 10^{12}$ L$_\odot$ year$^{-1}$;
Downes \& Solomon 1998), or $X = 0.8 ~ M_\odot ~ \rm (K~ km~ s^{-1}~
pc^{2})^{-1}$, and assuming constant brightness temperature from CO
5-4 down to CO 1-0 (ie. constant $L'(CO)$).  The implied gas mass is:
$M_{H_2} = 0.8 \times L'_{CO 1-0} = (1.6\pm 0.3) \times 10^{10}$
M$_\odot$. Note that the CO luminosity to gas mass conversion factor
is a factor five smaller for ULIRGs than it is for normal galaxies,
such as the Milky Way. For a detailed discussion of the conversion of
CO luminosity to gas mass in high redshift, FIR luminous galaxies, see
Solomon \& Vanden Bout (2005).

The ratio of FIR luminosity to CO luminosity has been used as a metric
for the relative star formation efficiency in galaxies
(ie. proportional to the star formation rate per unit gas mass; Gao \&
Solomon 2004). For J0927+2001, we find: $L_{FIR}/L'_{CO} = 650$
L$_\odot$ (K km s$^{-1}$ pc$^2$)$^{-1}$.  In their review of molecular
line emission from galaxies at $z > 1$, Solomon \& Vanden Bout (2005)
find that the values of $L_{FIR}/L'_{CO}$ range between roughly 100
and 1000, with a mean value of 350 L$_\odot$ (K km s$^{-1}$
pc$^2$)$^{-1}$, for galaxies with typical FIR luminosities between
10$^{12}$ and 10$^{13}$ L$_\odot$. For comparison, lower luminosity
star forming galaxies ($L_{FIR} < 10^{11}$ L$_\odot$) have
$L_{FIR}/L'_{CO}$ ratios an order of magnitude smaller (Kennicutt
1998).  J0927+2001 follows this general trend for increasing star
formation efficiency with increasing FIR luminosity.  These results
suggest a relatively short gas depletion timescale $\equiv$ (gas
mass)/(star formation rate) $\sim 10^7$ years, implying a brief, but
very intense, starburst.  This timescale is comparable to the typical
lifetime of luminous, high redshift quasars derived from the
clustering of quasars in the SDSS by Shen et al. (2007). Of course,
these arguments assume the FIR excess is due to dust heated by star
formation, and not the AGN. 

Although the gas mass in J0927+2001 is comparable to that in
J1148+5251, the line width of 600 km s$^{-1}$ is almost a factor two
broader than the CO emission from J1148+5251, and is at the high end of
the CO line width distribution observed for high redshift quasar host
galaxies (Carilli \& Wang 2006; Greve et al. 2005). A simple
explanation for a broader line would be that the gas disk is more
inclined to the sky plane in J0927+2001. High resolution imaging of
the CO distribution is required to determine the extent, and dynamics,
of the molecular gas. 

\section{Discussion}

J0927+2001 is the second example of an extreme molecular gas mass in a
galaxy within 1 Gyr of the Big Bang.  Like J1148+5251 at $z=6.42$,
J0927+2001 shows many characteristics of a co-eval massive starburst
in the host galaxy of the quasar (Section 1; Wang et al. 2007b). The
observations presented herein reveal the requisite molecular gas
reservoir to fuel the star formation. As in the case of J1148+5251,
Walter et al. (2004) point out that the formation of the heavy
elements, and in particular, the ISM processing required to form
cooler molecular gas and dust, implies that star formation must have
commenced very early in the host galaxy, at least a few hundred
million years prior, or $z > 8$.

These results lead to the question: how are such massive galaxies and
SMBH formed within 1 Gyr of the Big Bang?  Li et al. (2007a, b) have
addressed this question through multi-scale cosmological simulations,
including prescriptions for the complex processes of star formation
and AGN feed-back. They find that early galaxy and SMBH formation is
possible in rare (comoving density $\sim 1$ Gpc$^{-3}$), high density
peaks (halo mass $\sim 8\times 10^{12}$ M$_\odot$ at $z \sim 6$), in
the cosmic density field, through a series of gas-rich, massive
mergers starting at $z \sim 14$. SMBH formation occurs through a
combination of Eddington-limited accretion in each progenitor galaxy,
plus rapid black hole mergers during galaxy interactions.  The stellar
spheroids are formed in merger-driven, extreme starbursts, with star
formation rates exceeding 10$^3$ M$_\odot$ year$^{-1}$ for short
periods ($\sim 10^7$ years).  The ISM atomic abundances rapidly
approach Solar in the inner few kpc.  Li et al. (2007b) suggest that
such systems evolve into massive galaxies at the centers of the
densest cluster environments seen today ($\sim 10^{15}$ M$_\odot$).

Li et al. (2007a) hypothesize that systems such as J1148+5251 and
J0927+2001 may be in a late starburst phase, where the AGN has
recently emerged from its dusty shroud, and now potentially dominates
the dust heating, even for the warm dust component. The
question of the relative contribution of star formation and the AGN to
the heating of the warm dust that produces the observed (rest frame)
FIR excesses, will be discussed in detail in Wang et al. (2007b), and
can be addressed through the observations suggested below.

One outstanding issue the Li et al. models do not address is the early
formation of dust. Such early dust formation remains a puzzle, since
the standard ISM dust formation mechanism, ie. in the cool winds from
evolved low mass (AGB) stars, may require timescales longer than the
age of the universe at $z \sim 6$. One possible solution is dust
formation associated with massive star formation (Stratta et al. 2007;
Maiolino et al. 2004; Venkatesan, Nath, \& Shull 2006; Dwek et
al. 2007).

Such early coeval formation of galaxies and SMBHs has interesting
implications on the interpretation of the cosmic Stromgren spheres
around high redshift quasars, as inferred from the Gunn-Peterson
absorption spectra (Fan et al. 2006b). Masseli et al. (2007) and Lidz
et al. (2007) point out that, over the lifetime of the system, the
integrated star formation and AGN emission may contribute roughly
equally to the reionization of the regions immediately surrounding the
quasar host galaxy.

The detection of the gas reservoir required to fuel star formation in
the J0927+2001 quasar host galaxy is a key step in studying the
formation of massive galaxies at very high redshift, but clearly
further detailed observations are required to probe the host galaxy
and SMBH formation in this system. Critical diagnostic observations
include: imaging of the CO dynamics to constrain the galaxy
gravitational mass, and hence the black hole mass -- bulge mass
relation at high z (Shields et al. 2006; Walter et al. 2004),
multiwavelength imaging of the dust to help constrain dust heating
mechanisms, studies of lower order CO lines to determine gas
excitation and total gas mass (Bertoldi et al. 2004), detection of
higher density gas tracers, such as HCN, to determine the total mass
of gas directly associated with star forming clouds (Gao et al. 2007;
Krumholz \& Thompson 2007), and detection of the interstellar gas
cooling lines, such as [CII] (Maiolino et al. 2005; Iono et al. 2006).
The CO detection presented herein provides an accurate redshift for
the host galaxy as well as direct evidence for a highly developed ISM,
and hence is the first crucial step for future studies. For extreme
luminosity objects such as 0927+2001, many of these studies can be
performed with long integration times on current instruments, such as
the Plateau de Bure Interferometer and Very Large Array.  It will take
the Atacama Large Millimeter Array, with its more than an order of
magnitude increase in sensitivity relative to current millimeter
interferometers, to perform similar studies on less extreme systems,
such as the Ly-$\alpha$ galaxies at $z \sim 6$.

\acknowledgments 
The Plateau de Bure Interferometer is a facility of IRAM, supported by
INSU/CNRS (France), MPG (Germany), and IGN (Spain). CC, RW, JW
acknowledge support from the Max-Planck Society and the Alexander von
Humboldt Foundation through the Max-Planck Forschungspreise 2005.  The
National Radio Astronomy Observatory is a facility of the National
Science Foundation, operated by Associated Universities, Inc.
DR acknowledges support from the Deutsche Forschungsgemeinschaft (DFG)
Priority Program 1177.

\clearpage
\newpage

\begin{figure}[ht]
\includegraphics[width=6in,angle=-90]{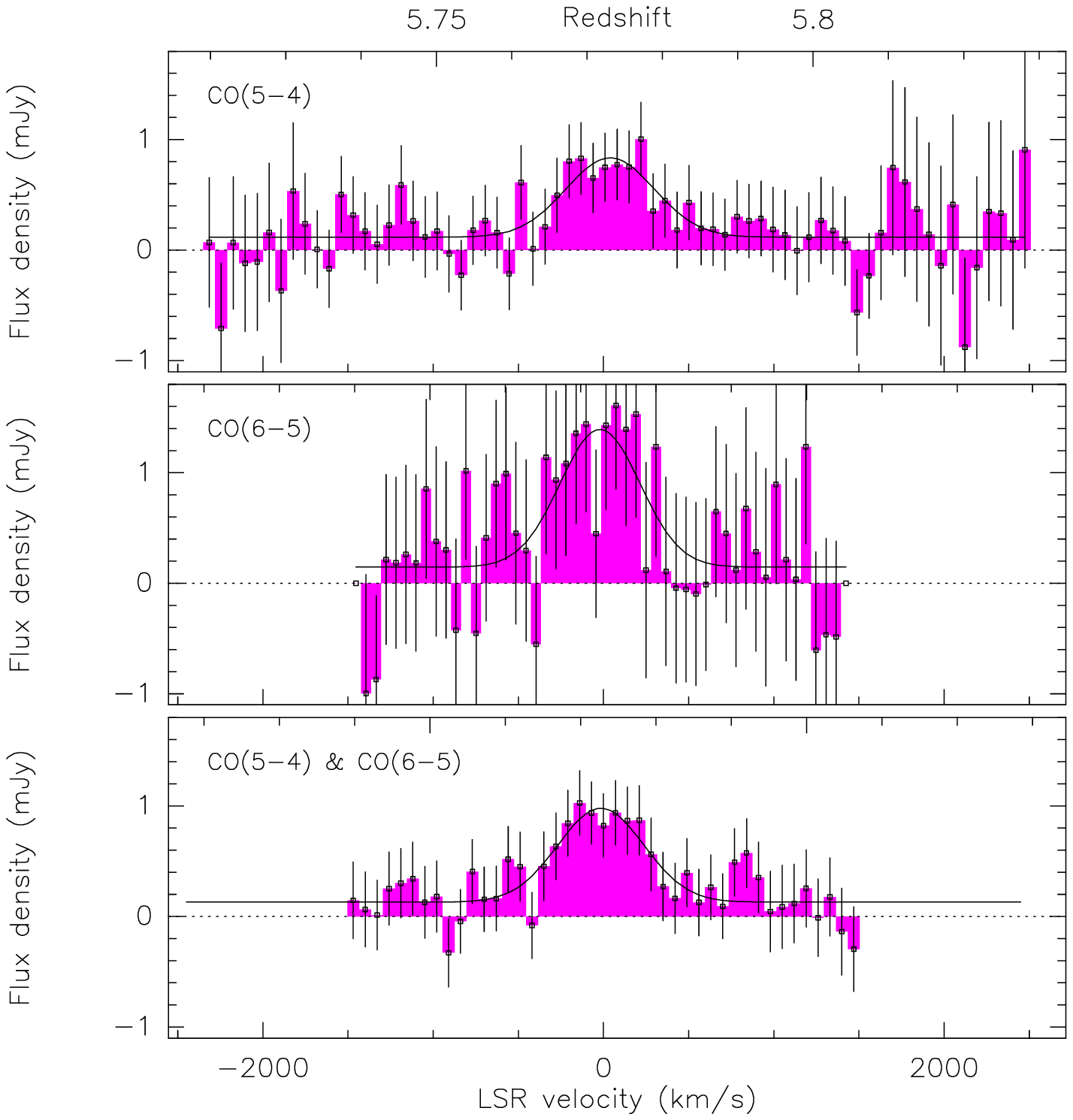}
\caption{The spectra of the CO 5-4 (upper), 6-5 (middle) and the sum
of the 5-4 and 6-5 (lower) emission from SDSS 
J0927+2001 at $z = 5.77$. Also shown
are Gaussian fits to the lines with parameters as given in Table 1.
All spectra are smoothed to 70 km s$^{-1}$ channel$^{-1}$, and each
independent channel is displayed. The rms
per channel varies across the band, although typical values in the
combined 5-4 and 6-5 spectrum are 0.3 mJy beam$^{-1}$.}
\label{}
\end{figure}

\clearpage

\begin{figure}[ht]
\includegraphics[width=4in,angle=-90]{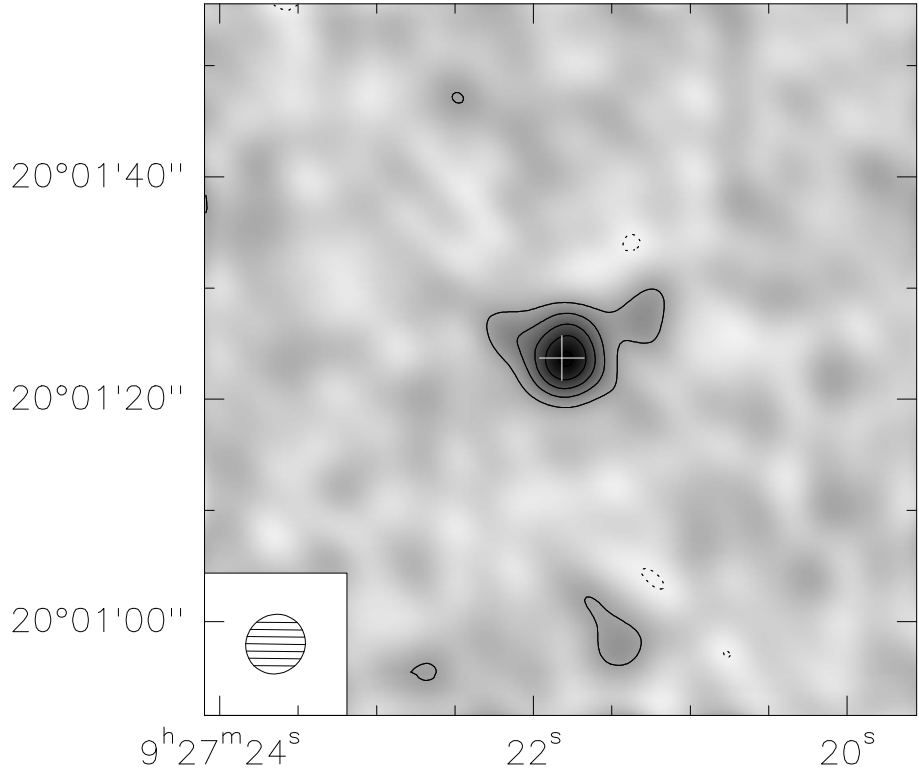}
\caption{An image of the sum of the velocity integrated CO 5-4 and CO
6-5 line emission from J0927+2001 at $z = 5.77$. Contour levels are:
0.14 Jy km s$^{-1}$ beam$^{-1}$.  The rms on the image is 0.07 Jy km
s$^{-1}$ beam$^{-1}$. The cross indicates the position of the optical
QSO. }
\label{}
\end{figure}

\clearpage

\begin{table}
\begin{center}
\caption{Gaussian fitting to the CO line emission from J0927+2001}
\begin{tabular}{cccccc}
\tableline\tableline
Transition & Continuum & Line Peak & FWHM & I$\Delta V$ & $L'(CO)$ \\
~ & mJy & mJy & km s$^{-1}$ & Jy km s$^{-1}$ & K km s$^{-1}$ pc$^{2}$ \\
\tableline
5-4 & $0.12\pm 0.03$ & $0.72\pm 0.09$ & $610\pm 110$ & $0.44\pm 0.07$ & $2.0\pm0.3 \times 10^{10}$ \\
6-5 & $0.13\pm 0.09$ & $1.25\pm 0.25$ & $550\pm 150$ & $0.69\pm 0.13$ & $2.2\pm0.5 \times 10^{10}$ \\
\tableline
\end{tabular}
\end{center}
\end{table}

\end{document}